# Ultrabright single-photon source on diamond with electrical pumping

Dmitry Yu. Fedyanin[1,*] and Mario Agio[2,3,4,5]

[1]*Laboratory of Nanooptics and Plasmonics, Moscow Institute of Physics and Technology, 141700 Dolgoprudny, Russian Federation*

[2]*National Institute of Optics (CNR-INO), 50125 Florence, Italy*

[3]*European Laboratory for Nonlinear Spectroscopy (LENS), 50019 Sesto Fiorentino, Italy*

[4]*Center for Quantum Science and Technology in Arcetri (QSTAR), 50125 Florence, Italy*

[5]*Laboratory of Nano-Optics, University of Siegen, 57072 Siegen, Germany*

*email: dmitry.fedyanin@phystech.edu

**The recently demonstrated electroluminescence of color centers in diamond makes them one of the best candidates for room temperature single-photon sources. However, the reported emission rates are far off what can be achieved by state-of-the-art electrically driven epitaxial quantum dots. Since the electroluminescence mechanism has not yet been elucidated, it is not clear to what extent the emission rate can be increased. Here we develop a theoretical framework to study single-photon emission from color centers in diamond under electrical pumping. The proposed model comprises electron and hole trapping and releasing, transitions between the ground and excited states of the color center as well as structural transformations of the center due to carrier trapping. It provides the possibility to predict both the photon emission rate and the wavelength of emitted photons. Self-consistent numerical simulations of the single-photon emitting diode based on the proposed model show that the photon emission rate can be as high as 100 kcounts s-1 at standard conditions. In contrast to most optoelectronic devices, the emission rate steadily increases with the device temperature achieving of more than 100 Mcount s-1 at 500 K, which is highly advantageous for practical applications. These results**

**demonstrate the potential of color centers in diamond as electrically driven non-classical light emitters and provide a foundation for the design and development of single-photon sources for optical quantum computation and quantum communication networks operating at room and higher temperatures.**

**Introduction**

Operation at the single-photon level promises the ultimate energy efficiency for optical and optoelectronic devices and opens new prospects for various applications, such as secure optical communications based on quantum cryptography [1] and quantum computers [2]. In this respect, single-photon sources operating upon electrical injection are vitally important [3–5], since they are characterized by higher energy efficiency and better integration compatibility compared to optically pumped devices.

Epitaxial quantum dots embedded into a p-i-n diode structure were proposed to be used as electrically driven single emitters [3,6–8], however their operation is limited to cryogenic temperatures. Apart from this, spontaneous emission control in this structure is complicated [9,10]. This drawback can be eliminated by utilizing nanowire heterostructures [10,11]. Such an approach allows employing cavity quantum electrodynamic (QED) effects to achieve high extraction efficiency [12,13]. Nevertheless, operation is still limited to low temperatures. Room temperature operation is highly desirable, being the case, where the net energy efficiency of the devices is high enough for integration in practical optical and photonic systems.

At present, one of the best candidates for a stable single-photon source operating at room temperature are color centers in diamond [14,15]. These defects in the lattice structure have been studied intensively during the last two decades and optically driven single-photon sources have been demonstrated, but the possibility of operation upon electrical injection was not clear until very recently, mainly because diamond is a unique material at the interface between solid-state and semiconductor physics and can demonstrate the effects from both of these worlds [16,17]. A great progress in diamond electronics has enabled the injection electrons and holes into defects in diamond, and the demonstrated electroluminescence from nitrogen-vacancy color centers [18,19] has proven the feasibility of electrically driven single-photon sources on diamond. However, the luminescence intensity of the order of only $10^4$ counts s$^{-1}$ was observed [19], which is much lower than that of the optically pumped color centers [20,21] and electrically pumped semiconductor structures [22]. Without the theoretical description of electrically pumped color centers in diamond, it is not clear to what extent the photon emission rate can be increased.

Here, we introduce a model that qualitatively and quantitatively describes single-photon emission from color centers in diamond under electrical pumping. It considers electron and hole trapping and releasing, transitions between the ground and excited levels of the center as well as structural transformations of the center due to carrier trapping. The latter are reflected in the energy-level structure of the color center, which determines its emission properties. Using the developed model, we predict the photon emission rate from electrically pumped color centers and explain the difference in their electroluminescence, which is observed from either a charged or neutral charge state for different color centers. Furthermore, using a comprehensive computational approach, we study electroluminescence of a single color center embedded in a p-n junction diamond diode in a wide temperature range and demonstrate that, in contrast to most semiconductor and solid-state devices, the diamond single-photon emitting diode exhibits superior emission properties at high temperatures with emission rates exceeding $10^8$ counts per second.

**Results**

**Recombination at the color center and photon emission**

The photoluminescence of color centers in diamond has been broadly studied for years [21,23]. However, electroluminescence and photoluminescence processes are remarkably different from each other. Figure 1(a) illustrates the basic principle of photon emission under optical pumping: an intense external light source pumps the color center from the electronic ground state to some state above the lowest excited level. This short-lived state quickly relaxes to the emitting state and the color center then returns to the ground state emitting one photon with energy $\hbar\omega = E_{|e\rangle} - E_{|g\rangle}$. Thus, such a photoluminescence process involves only transitions between the ground and excited states of the color center and excludes its interaction with the conduction and valence bands of diamond, which allows to consider the color center as an isolated atom-like system. Electroluminescence, in turn, is a much more complicated process, which essentially implies interaction between the color center and diamond conduction and valence bands via recombination processes.

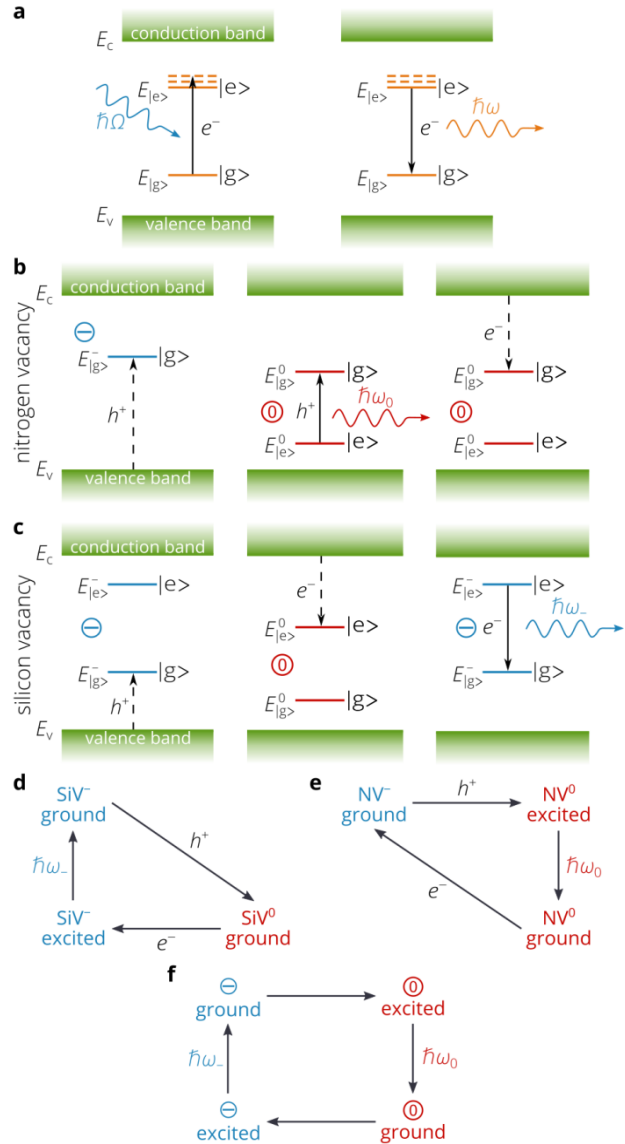

**Figure 1: Mechanism of photon emission from color centers in diamond under electrical pumping.** (a) Schematic diagram of single-photon emission from a color center in diamond under optical pumping. The color center makes an optical transition from the electronic ground state to an excited level, which typically quickly relaxes to the lowest excited state. A photon is emitted and the color center returns to the ground state. (b) Schematic illustration of the three-stage process of single-photon emission upon electrical injection from the NV-center in diamond. In the first stage, a hole is trapped by the negatively charged center to the excited level. Simultaneously, hole trapping alters the charge of the NV-center and consequently changes its energy level structure. Thus, the trapped hole appears at the excited

level for holes in the neutral center, which, in fact, is the well-known excited state of the $NV^0$-center. In the second stage, the excited state relaxes to the ground state via photon emission. Finally, the NV-center returns to the initial negative charge state capturing an electron. The sign in a circle and color indicate the charge state of the color center: ⊖ and blue color correspond to the negatively charged center and ⓪ and red color to the neutral one. (c) Schematic representation of the three-stage process of single-photon emission upon electrical injection from the SiV-center in diamond. (d, e) Diagrams illustrating the electroluminescence process of the SiV-center (d) and NV-center (e). (f) Schematic diagram of electroluminescence of a hypothetical center with two (negative and neutral) charge states, each of which has excited states, which can be excited electrically.

Let us consider first an electrically pumped nitrogen vacancy (NV) complex, which is the most studied color center in diamond. This acceptor type defect in the crystal lattice has two charge states: negatively-charged ($NV^-$) and neutral ($NV^0$), both exhibiting optically induced radiative transitions between ground and excited states [24]. Limiting our study to a simplified two-level model, photon emission can be treated as a three stage process [Fig. 1(b,c)]. We further assume the color center to be negatively charged in equilibrium. This assumption does not affect the generality of our model, since the studied process is cyclic and the color center returns to the initial state at the end of each cycle passing through both charge states, as seen below. A negatively-charged defect is an attractive center for holes. It has a large capture cross-section $\sigma_p$ of up to $10^{-12}$ cm$^2$ due to the low mobility of free carriers in heavily doped diamond [17]. The $NV^-$ center accepts a hole from the valence band of diamond [Fig. 1(b)], which alters the charge of the defect and consequently changes its energy level structure. At the same time, the defect traps a hole into the excited level [25], which is the excited level for holes [26] in the neutral center (since the trapped hole has changed the defect charge) [Fig. 1(b)]. Such an excited level for holes can be viewed as an energy level $E^0_{|e\rangle}$ in the bandgap of diamond between the ground level $E^0_{|g\rangle}$ of the NV$^0$ center and the valence band edge of diamond $E_v$. This level actually corresponds to the $^3E$ excited state of the NV$^0$ center [23]. Next, the excited state of the NV$^0$ center relaxes to ground state, which can be viewed as a radiative transition between the $E^0_{|e\rangle}$ and $E^0_{|g\rangle}$ levels [Fig. 1(b)]. Therefore, the wavelength of the emitted photon coincides with that of the optically excited NV$^0$ center. Since the color center has only two charge states (negative and neutral), the neutral center cannot take up holes anymore, but it can capture electrons with a capture cross-section $\sigma_n$ of about $10^{-15}$ cm$^2$ [27–29]. Thereby, the color center returns to the initial negative charge state [Fig. 1(b)]. The proposed model is in a good agreement with the experimental measurements in Ref. [19], where electroluminescence only from neutral NV$^0$ centers was observed. In spite of the hypothesis that electroluminescence from negatively charged NV$^-$ centers can be also obtained [19], a detailed analysis of the energy level structure of the NV$^-$ center [30], which is a multielectron system, demonstrates that an electron cannot be trapped into the excited state of the NV$^-$ center, since the energy received by the center from electron trapping is lower than the energy required to bring the center from the NV$^0$ ground state to the NV$^-$ excited state.

In general, emission properties of color centers are determined by their energy level structure. Hence, some color centers with two charge states (neutral and charged) can exhibit electroluminescence from the charged state, but, at the same time, their neutral states will not show any photon emission under electrical injection [Fig. 1(d)]. The silicon vacancy (SiV) defect in diamond is a good example for such a color center [31]. In contrast to the NV center, it cannot capture a hole into the SiV$^0$ excited state because there is only one electron level above the valence band edge in the SiV$^-$ ground states, which is occupied by three electrons [32]. Capturing a hole in this level (i.e. releasing an electron to the valence band of diamond) immediately transforms the center into the SiV⁰ ground states. It is interesting that some color centers can in principle demonstrate quite unique properties emitting serially two photons at different wavelengths under electrical pumping [Fig. 1(e)], while others may not show electroluminescence. The present model can be easily extended to a multicharged center, which has three and more charge states. However, since the electron (hole) capture cross-section by the repulsive negatively (positively) charged center is as low as $10^{-22} - 10^{-19}$ cm² [25], only single charged and neutral charge states can achieve high photon emission rates under electrical pumping.

Considering a detailed balance between the radiative transition from the excited level to the ground level, electron and hole trapping from the conduction and valence bands of diamond to the levels of the NV center and carrier thermal escape from the local levels back to the diamond bands, we obtain the photon emission rate $R_{\mathrm{ph}}$ in the steady state as (for details, see Supplementary Information)

$$R_{\mathrm{ph}} = \Phi c_{\mathrm{n}} n \frac{\left(1 + \dfrac{p_1}{p} + \dfrac{n_1 p_1}{np}\right) f - \dfrac{n_1 p_1}{np}}{1 + \dfrac{p_1}{p} + \dfrac{c_{\mathrm{n}} n_1}{c_{\mathrm{p}} p}}. \quad (1)$$

Here, $\Phi$ is the quantum efficiency, $n$ and $p$ are respectively the electron and hole densities in diamond, $c_{\mathrm{n}} = \sigma_{\mathrm{n}} \langle \upsilon_{\mathrm{n}} \rangle$ and $c_{\mathrm{p}} = \sigma_{\mathrm{p}} \langle \upsilon_{\mathrm{p}} \rangle$ are the electron and hole capture rate constants, $\langle \upsilon_{\mathrm{n}} \rangle = \left(8 k_{\mathrm{B}} T / \pi m_{\mathrm{n}}^{\mathrm{cond}}\right)^{1/2}$ and $\langle \upsilon_{\mathrm{p}} \rangle = \left(8 k_{\mathrm{B}} T / \pi m_{\mathrm{p}}\right)^{1/2}$ are the average carrier thermal velocities, $m_{\mathrm{p}}$ is the effective hole mass and $m_{\mathrm{n}}^{\mathrm{cond}}$ is the electron conductivity effective mass. Finally, the variables $n_1$ and $p_1$ are similar to the notations used in the Shockley-Read-Hall recombination model [33] and have the meaning of the equilibrium electron ($n_1$) and hole ($p_1$) densities when the Fermi level in diamond coincides with the ground level $E^-_{|g\rangle}$ of the NV$^-$ center and excited level $E^0_{|e\rangle}$ of the NV$^0$ center, respectively. Finally, $f$ is the occupation probability of the neutral ground state given by

$$f = \frac{1}{c_n n} \frac{1 + \dfrac{c_n n_1}{c_p p} + \dfrac{p_1}{p} \dfrac{c_n n_1}{1/\tau}}{\dfrac{1}{c_n n} + \dfrac{1}{c_p p}\left(1 + \dfrac{n_1}{n}\right) + \tau\left(1 + \dfrac{p_1}{p} + \dfrac{n_1 p_1}{np}\right)}, \qquad (2)$$

where $\tau$ is the excited-state lifetime. For electrons, the capture cross-section of the center in the neutral state is about the lattice constant of diamond and is estimated to be $10^{-15}$ cm$^2$ [27–29] and, accordingly, $c_n \approx 1.7\times10^{-8}$ cm$^3$s$^{-1}$. At the same time, the color center in the charge state is the attractive center for holes, which can increase the capture cross-section by orders of magnitude. Based on the cascade capture model [34], $c_p$ is evaluated as $4\pi r_T^3/3\tau_E = 3.9\times10^{-7}$ cm$^3$s$^{-1}$, where $r_T$ is the critical radius in the Thomson model [35] and $\tau_E$ is the relaxation time of the hole energy due to interaction with acoustic phonons [36].

A brief analysis of equations (1) and (2) shows that, if the ground and excited levels of the color center are located deep in the bandgap of diamond, the effective densities $n_1$ and $p_1$, which characterize thermal emission rates from the local levels to the conduction and valence bands (see Supplementary Information), are reduced to zero and the recombination rate at the color center (i.e., the photon emission rate) $R_{ph}$ is defined simply as the inverse of the sum of the three inverse rate: electron capture ($c_n n$), hole capture ($c_p p$) and transition rate from the exited to ground level ($1/\tau$):

$$R_{ph} = \Phi \frac{1}{\dfrac{1}{c_n n} + \dfrac{1}{c_p p} + \tau}. \qquad (3)$$

Since $\tau$ is of the order of a few nanoseconds [21], $R_{ph}$ is mostly limited by slower trapping processes. These require the concentrations of both carriers to exceed $10^{15}$ cm$^{-3}$ in order to be as fast at the transition from the excited to the ground level of the color center. However, such high free carrier concentrations are hardly achievable at room temperature due to the high ionization energies of donors and acceptors [37,38] and strong compensation effects in diamond [37–40], which are especially pronounced for electrons [37,39,40]. In addition, non-equilibrium carrier concentrations in the active region of the semiconductor device operating under electrical injection are typically lower than the equilibrium majority-carrier concentrations in the bulk [41]. Therefore, in order to understand to what extent the single-photon emission can be increased, we proceed to the analysis of the diamond p-n diode with a color center in the vicinity of the junction.

**Single-photon-emitting diamond diode**

Figure 2(a) shows a schematic of the p-n junction structure with an embedded color center. The p-type layer is uniformly doped with boron. The acceptor concentration is equal to $N_A = 5 \times 10^{18}$ cm$^{-3}$ and the compensation ratio $\eta_p$ is as small as 1%, which is typical for p-type diamond samples [38,41] and corresponds to a hole density of $3.9 \times 10^{14}$ cm$^{-3}$. Under similar doping conditions ($N_D = 3 \times 10^{18}$ cm$^{-3}$), the electron density in the n-type layer is much lower ($6.3 \times 10^{10}$ cm$^{-3}$) due to stronger compensation effects ($\eta_n = 10\%$ [42–44]). In spite of very small carrier densities in the bulk, donor and acceptor are almost fully ionized near the p-n junction and the thicknesses of the depletion layers, where the bands are bent, do not exceed 30 nm at both sides of the junction, which makes the present electrically driven single photon source truly nanoscale. The electron and hole mobilities are limited by carrier scattering on ionized and nonionized impurities and are calculated to be about $\mu_n = 220$ cm$^2$V$^{-1}$s$^{-1}$ and $\mu_p = 290$ cm$^2$V$^{-1}$s$^{-1}$ throughout the n-type and p-type regions [16,36] and the carrier lifetime determined by the Shockley-Read-Hall recombination is estimated to be 50 ps. Using a self-consistent steady-state model, which comprises the Poisson equation, the drift-diffusion current equations and the electron and hole continuity equations to describe the carrier behavior in the diamond n-type and p-type layers [45,46], we simulate the electron and hole transport in the 600 nm long diode shown in Fig. 2(a) [Fig. 2(b–d)]. Under forward bias, holes are injected from the p-side into the n-side and electrons from the n-side into the p-side. These non-equilibrium carriers can be captured by the color center, which results in light emission. The densities of injected minority carrier decay exponentially with the distance from the junction and, consequently, the photon emission rate strongly depends on the position of the color center with respect to the p-n junction, since $R_{ph}$ is proportional to the carrier densities as follows from equations (1) and (2). Figure 2(e) reports the maximum photon emission rate, which can be achieved at each injection current (bias voltage).

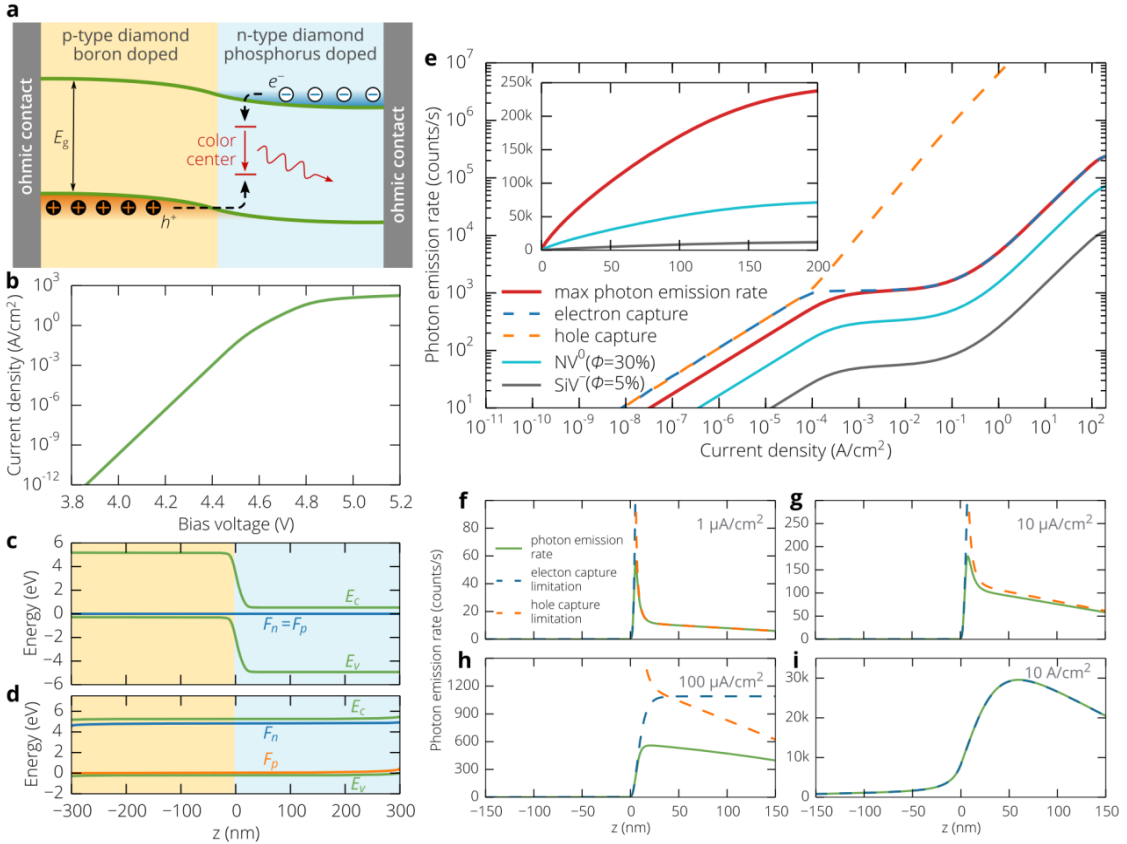

**Figure 2: Single-photon emitting diamond diode.** (a) Schematic of the p-n diamond diode with a color center implanted in the vicinity of the p-n junction. The p-type and n-type

regions have the same thickness of 300 nm. (b) Simulated current-voltage characteristic of the diode shown in panel (a). (c–d) Energy band diagram in equilibrium (c) and under high forward bias of 4.9 eV (d). (e) Dependence of the maximum single-photon emission rate on the injection current density for a color center with deep energy levels, which has the neutral and negative charge states. The dashed lines show the contribution of the electron capture ($c_n n$) and hole capture ($c_p p$) process to the recombination at the color center given by equations (1) and (2) in the optimal position of the color center, which corresponds to the maximum emission rate for each bias voltage. Also shown are the emission rates from the $NV^0$ and $SiV^-$ centers, which are characterized by the quantum efficiencies of $\Phi = 30\%$ [47] and $\Phi = 5\%$ [21], respectively. (f-i) Photon emission rate of the color center for 100% quantum efficiency as a function of its position with respect to the p-n junction at different injection currents: (f) – 1 µA cm$^{-2}$, (g) – 10 µA cm$^{-2}$, (h) – 100 µA cm$^{-2}$ and (i) – 10 A cm$^{-2}$.

In the low injection current regime ($V < 4.5$ V or $J < 100$ mA cm$^{-2}$), the current-voltage characteristic exhibits a clear exponential dependence [Fig. 2(b)] typical for conventional semiconductor p-n diodes [41]. At very small current densities, the densities of the injected carriers decrease rapidly with distance from the junction [Fig. 2(f,g)] and the maximum emission rate is observed for the color center placed at a distance $z = 3 - 8$ nm, where electron and hole trapping processes contribute equally to $R_{ph}$, which is clearly seen in Fig. 2(e). The slight deviation of the optimal position of the color center from $z = 0$ is determined by the high difference in the majority carrier concentration in the n-type and p-type layers and capture cross-section for electrons and holes. As the bias voltage further increases, the hole capture rate rapidly increases and becomes greater than the electron capture rate all through the n-type layer [Fig. 2(h)] which is clearly shown in Fig. 2(e), where the contribution of the electron and hole capture processes are plotted. Recombination at the color center becomes limited by the electron trapping $c_n n_{bulk}$ = 1.1 kcounts s$^{-1}$ and the subsequent increase of the injection current does not increase the photon emission rate until the diode reaches the high injection current regime.

At high injection levels $J >$ 100 mA cm$^{-2}$, the bias voltage is so high that it produces band bending in the n-type and p-type regions [see Fig. 2(d)], which in turn affects the majority carrier spatial distributions. This effect gives a possibility for the color center to emit more than 1.1 kcounts s$^{-1}$, limited by the equilibrium electron density in the n-type layer [Fig. 2(i)]. This process is accompanied by the shift of the optimal position of the color center in the bulk of the n-type layer. Eventually, the photon emission rate can exceed 200 kcounts s$^{-1}$ at very high injection currents (~100 A cm$^{-2}$), which, however, produce heating of the diamond diode and consequently proper temperature stabilization is required to maintain the temperature at 300 K.

Here we should note that, at moderate and high injection levels, the recombination rate at the color center is limited by the electron density [see Fig. 2(e)], which is known to be very small in n-type diamond due to the high activation energy of donors and strong compensation effects [37,39,43,44]. However, as temperature increases, the probability of electron thermal emission from the deep donor to the conduction band increases. Therefore, the density of free electrons in the n-type layer grows rapidly with temperature as $T^{3/2}\exp(-E_D/k_B T)$ (see insert in Fig. 3), so does the photon emission rate from the color center. Fig. 3 demonstrates that a small temperature increase from 300 K to 350 K can improve the photon emission rate by more than one order of magnitude and, at 500 K, it becomes possible to achieve 10 Mcounts s$^{-1}$. Thus, instead of cooling the device to room or lower temperatures, which usually improves the performance of semiconductor devices [41] (and single photon sources in particular [8−11]), one can improve the emission properties of the electrically pumped color center by heating the device. Undeniably, the quantum yield slightly decreases with temperature [23,48], but this effect can be compensated by using optical antennas or cavities [49–51]. Furthermore, the temperatures of about 200 °C can easily be achieved by using conventional dielectrics with a low thermal conductivity (e.g. $SiO_2$ or $HfO_2$) and self-heating effects. Hence, the electrically driven single-photon source can operate at room temperature in an ambient environment without any integrated heaters.

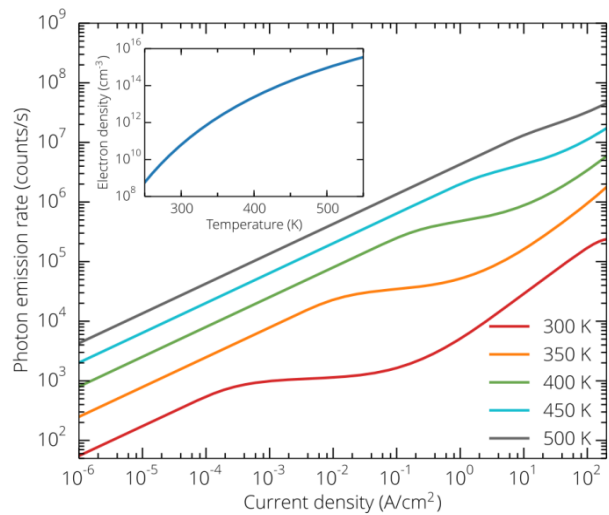

**Figure 3: Input-output characteristics of the single-photon emitting diamond diode above room temperature.** Maximum single-photon emission rate from the color center with deep energy levels, which has neutral and negative charge states, versus the injection current density at different lattice temperatures. Insert: Electron density in the n-type layer of the diamond p-n diode as a function of temperature, $\eta_n = 10\%$ and $E_D = 0.57$ eV.

A bright emission at room temperature cannot be achieved due to strong donor compensation effects and high donor activation energy. At a donor compensation ratio $\eta_n$ of 10%, the photon emission rate saturates at a level of about $10^3$ counts s$^{-1}$ until strong band bending due to high bias voltage produces a well for electrons in the n-type region increasing the electron density nearby the color center. However, strong donor compensation in diamond is mainly a technological problem [52] and it was already experimentally demonstrated that $\eta_n$ can be reduced to 0.4%, which provides much higher electron densities in diamond and, consequently, up to more than one order of magnitude increase in the photon emission rate (Fig. 4). At the same time, at higher temperatures ($\gtrsim$ 400 K), the difference in the output power between the single-photon emitting diodes with different compensation ratios of the n-type region does not exceed 4-fold, which demonstrates the robustness of these devices with respect to technological limitations. Nevertheless, Fig. 4 demonstrates that better diamond samples should provide higher emission rates and further technological advancements in diamond electronics can potentially move the operation speed of electrically driven single-photon source to the gigahertz range, which is significantly higher than the actual bandwidth (1 – 10 Mbit s$^{-1}$) of the quantum key distribution systems based on attenuated lasers [53].

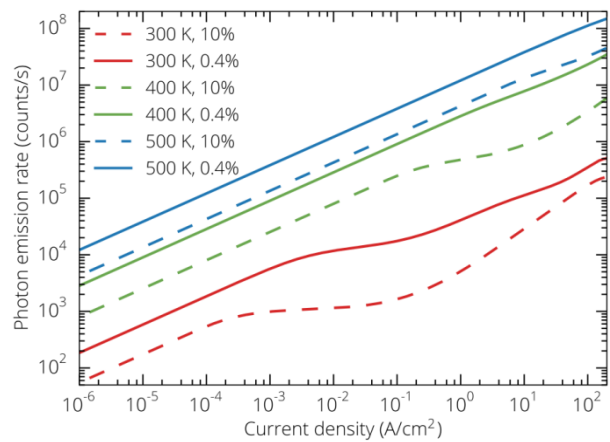

**Figure 4: Input-output characteristics at moderate and low compensation ratios of the n-region of the diamond diode.** Single-photon emission rate from the electrically pumped color center in the p-n diode configuration [see Fig. 2(a)] versus the injection current density at different operation temperatures and compensation ratios of the n-region of the diamond diode.

## Discussion

We have established a theoretical framework to study single photon emission from color centers in diamond upon electrical injection. The introduced model is in good quantitative agreement with the experimental results and shows that the process of electroluminescence is remarkably different from that of photoluminescence. Whereas photoluminescence involves only transitions between the ground and excited states of the color center, electroluminescence comprises electron and hole trapping, transitions between the ground and excited levels of the center as well as structural transformations of the center due to electron/hole trapping and releasing. We have shown that, due to these transformations, photon emission can be observed only from the neutral charge state of the NV center and negative charge state of the SiV center. This is determined by the energy-level structures of the color centers, which are actually multielectron systems and this cannot be ignored, when considering electron and hole capture processes. In addition, we have shown that, in principle, some color centers may exhibit remarkable emission properties under electrical pumping, namely the serial emission of two photons at two different wavelengths: the energy of the first photon corresponds to the transition between the excited and ground states of the neutral (charged) center and the energy of the second one to that of the charged (neutral) center. Finally, we should note that our theory can be easily applied to study electrically pumped color centers in silicon carbide [54], for which manufacturing technologies are well developed.

We have also rigorously investigated an electrically driven color center in the p-n diode configuration and found that the photon emission rate is mostly limited by slow electron and hole capture processes, which, nevertheless, can be relatively fast at high injection levels giving a possibility to achieve a single-photon photon emission rate of the order of $10^5$ counts per second at room temperature. Further increase of the emission rate is seriously suppressed by the high activation energy of donors in diamond and compensation effects, since the capture rates are determined not only by the capture cross-sections but also by the free carrier densities, which in turn strongly depend on the dopant concentrations and their activation energies. At the same time, the activation energy of phosphorus, which is known to date to be the only dopant in diamond that can provide high electron concentrations [55], is as large as 0.6 eV. Accordingly, even a barely perceptible density of compensating defects and impurities dramatically reduce the density of free electrons. In spite of these difficulties, we have shown that, in contrast to most semiconductor optoelectronic devices, the emission properties can be significantly improved by heating the diamond diode to 100 – 200 °C. This gives a possibility to increase the photon emission rate up to $10^8$ counts per second. Further advances in diamond doping easily move the bandwidth of diamond single-photon emitting diodes to the gigahertz range. These results create the backbone for the design and development of practical electrically driven single-photon sources operating at room or higher temperatures.

## Acknowledgements

The work was supported by the Russian Science Foundation (14-19-01788) and the EC Seventh Framework Programme (248855).

# Supplementary Information for

# Ultrabright single-photon source on diamond with electrical pumping

Dmitry Yu. Fedyanin[1,*] and Mario Agio[2,3,4,5]

[1]Laboratory of Nanooptics and Plasmonics, Moscow Institute of Physics and Technology, 141700 Dolgoprudny, Russian Federation

[2]National Institute of Optics (CNR-INO), 50125 Florence, Italy

[3]European Laboratory for Nonlinear Spectroscopy (LENS), 50019 Sesto Fiorentino, Italy

[4]Center for Quantum Science and Technology in Arcetri (QSTAR), 50125 Florence, Italy

[5]Laboratory of Nano-Optics, University of Siegen, 57072 Siegen, Germany

*email: dmitry.fedyanin@phystech.edu

**Electroluminescence model**

Consider a defect in the crystal lattice of diamond, which has two charges states (neutral and negative) and the neutral charge state has an excited level for holes (Fig. S1). As shown in the main text, this simplified model can be used to describe photon emission from electrically pumped NV centers in diamond. The recombination process at the defect resulting in photon emission can be described as follows: a hole is captured by the negatively charged center to the excited state of the neutral center and, after that, this excited state relaxes to the ground state via photon emission. At the same time, an electron is captured by the neutral center to the ground state of the negatively charged center [Fig. S1(a)].

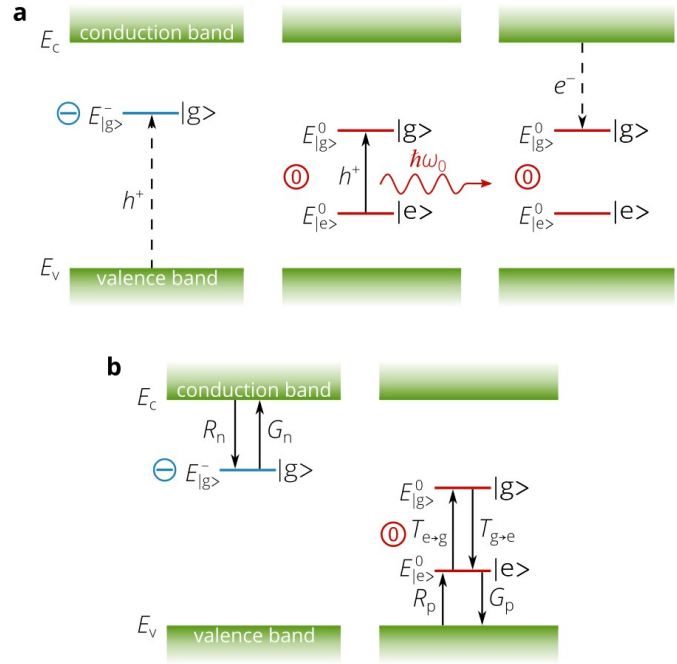

**Figure S1. (a)** Diagram of the three-stage process of single-photon emission upon electrical injection from the color center. **(b)** Schematic illustration of electron and hole transitions involved in recombination at the color center, the transition rates are also marked. In both panels, the sign in a circle and color indicate the charge state of the color center: ⊖ and blue color correspond to the negatively charged center and ⓪ and red color to the neutral one.

Electron and hole transitions among the conduction band of diamond, levels of the color center and valence band of diamond involved in the recombination process are shown schematically in Fig. S1(b). The capture rate of electrons by the color center per unit volume is given by [1]

$$R_n = c_n\, n\, f\, N. \tag{S1}$$

Here, $c_n$ is the electron capture rate constant, $n$ is the electron concentration in diamond, $N$ is the number of color centers per unit volume and $f$ is the population of the neutral ground state. Similarly, the hole capture rate per unit volume $R_h$ is proportional to the density of holes $p$ and the number of centers in the negative charge state per unit volume $(1 - x - f)N$:

$$R_p = c_p\, p\, (1 - x - f)\, N, \tag{S2}$$

where $c_p$ is hole capture rate constant and $x$ is the population the neutral excited state. Electron and hole thermal emission processes can be expressed in the same manner:

$$G_n = e_n(1-x-f)N, \tag{S3}$$

$$G_p = e_p x N, \tag{S4}$$

where $e_n$ and $e_p$ are the thermal emission constants for electrons and holes, respectively. The hole transition rate from the excited level to the ground level can be characterized by the lifetime $\tau$, which includes both radiative and non-radiative processes:

$$T_{e \to g} = \frac{1}{\tau} x N. \tag{S5}$$

Here, we assume the quantum efficiency of this transition to be equal to $\Phi$. Accordingly, the photon emission rate from the color center under electrical pumping can be obtained simply by multiplying the recombination rate at the center by $\Phi$. The rate of the inverse process (hole transition from the ground level to the excited level) is expressed as

$$T_{g \to e} = e_r f N. \tag{S6}$$

In Equation (S6), $e_r$ is the constant that characterizes thermal excitation from the neutral ground state to the neutral excited state.

Following Abakumov and Rzhanov [2,3], we can write balance equations for electrons and holes in diamond as well as for the populations of the ground and excited states:

$$\begin{cases} \dfrac{dn}{dt} = G_n - R_n + G_n^{ext} = e_n(1-x-f)N - c_n n f N + G_n^{ext} \\ \dfrac{dp}{dt} = G_p - R_p + G_p^{ext} = e_p x N - c_p p(1-x-f)N + G_p^{ext} \\ \dfrac{d(fN)}{dt} = G_n - R_n + T_{e \to g} - T_{g \to e} = e_n(1-x-f)N - c_n n f N + xN/\tau - e_r f N \\ \dfrac{d(xN)}{dt} = R_n - G_n + T_{g \to e} - T_{e \to g} = c_p p(1-x-f)N - e_p x N + e_r f N - xN/\tau \end{cases} \tag{S7}$$

Here, $G_n^{ext}$ and $G_p^{ext}$ are the electron and hole generation rates due to electrical pumping. In equilibrium, the probability to find the color center in each of three (negative ground, neutral ground and neutral excited) states can be easily found using Gibbs' distribution [4]:

$$f_0 = \frac{g^0_{|g\rangle} \exp\left(\frac{E^0_{|g\rangle} - F}{k_B T}\right)}{g^-_{|g\rangle} + g^0_{|g\rangle} \exp\left(\frac{E^0_{|g\rangle} - F}{k_B T}\right) + g^0_{|e\rangle} \exp\left(\frac{E^0_{|e\rangle} - F}{k_B T}\right)} \tag{S8}$$

and

$$x_0 = f_0 \frac{g^0_{|e\rangle}}{g^0_{|g\rangle}} \exp\left(\frac{E^0_{|e\rangle} - E^0_{|g\rangle}}{k_B T}\right), \tag{S9}$$

where $g^-_{|g\rangle}$, $g^0_{|g\rangle}$ and $g^0_{|e\rangle}$ are the degeneracies of the negative ground, neutral ground and neutral excited states, respectively, and $F$ is the Fermi level in diamond. Since at equilibrium $dn/dt = dp/dt = 0$, $d(fN)/dt = 0$, $d(xN)/dt = 0$ and $G^{ext}_n = G^{ext}_p = 0$, one easily finds that

$$e_n = c_n n_1, \tag{S10}$$

$$e_p = c_p p_1, \tag{S11}$$

$$e_r = \frac{1}{\tau} \frac{g^0_{|e\rangle}}{g^0_{|g\rangle}} \exp\left(\frac{E^0_{|e\rangle} - E^0_{|g\rangle}}{k_B T}\right), \tag{S12}$$

where

$$n_1 = \frac{g^0_{|g\rangle}}{g^-_{|g\rangle}} N_c \exp\left(\frac{E^0_{|g\rangle} - E_c}{k_B T}\right), \tag{S13}$$

and

$$p_1 = \frac{g^-_{|g\rangle}}{g^0_{|e\rangle}} N_v \exp\left(\frac{E_v - E^0_{|e\rangle}}{k_B T}\right). \tag{S14}$$

In the above expressions, $N_c$ and $N_v$ are the effective densities of states in the conduction and valence bands of diamond, respectively.

Further, we can find stationary populations for the states of the color center. In steady state, all time derivatives in the system of Equations (S7) are equal to zero and $R_n - G_n = R_p - G_p$. Solving the balance equations under these conditions, after a few algebraic steps we obtain the probability to find the color center in the neutral ground state:

$$f_{ss} = \frac{1}{c_n n} \frac{1 + \frac{c_n}{c_p} \frac{n_1}{p} + \frac{c_n n_1}{1/\tau} \frac{p_1}{p}}{\frac{1}{c_n n}\left(1 + \frac{e_r}{1/\tau}\right) + \frac{1}{c_p p}\left[1 + \frac{e_r}{1/\tau}\left(1 + \frac{p_1}{p}\right) + \frac{n_1}{n}\right] + \tau\left[1 + \frac{p_1}{p}\left(1 + \frac{n_1}{n}\right)\right]}. \quad (S15)$$

Since $\left(E^0_{|g>} - E^0_{|e>}\right) \gg kT$, Equation (S15) can be reduced to Equation (2) in the main text:

$$f_{ss} \approx \frac{1}{c_n n} \frac{1 + \frac{c_n}{c_p} \frac{n_1}{p} + \frac{c_n n_1}{1/\tau} \frac{p_1}{p}}{\frac{1}{c_n n} + \frac{1}{c_p p}\left(1 + \frac{n_1}{n}\right) + \tau\left[1 + \frac{p_1}{p}\left(1 + \frac{n_1}{n}\right)\right]}. \quad (S16)$$

Collecting the above results, we get the stationary recombination rate at a single color center

$$R_{SCS} = c_n n \frac{f_{ss}\left(1 + \frac{p_1}{p} + \frac{p_1 n_1}{pn}\right) - \frac{p_1 n_1}{pn}}{1 + \frac{c_n n_1}{c_p p} + \frac{p_1}{p}}. \quad (S17)$$

Finally, the photon emission rate from this center is given by

$$R_{ph} = \Phi R_{SCS} = \Phi c_n n \frac{f_{ss}\left(1 + \frac{p_1}{p} + \frac{p_1 n_1}{pn}\right) - \frac{p_1 n_1}{pn}}{1 + \frac{c_n n_1}{c_p p} + \frac{p_1}{p}}, \quad (S18)$$

which corresponds to Equation (1) in the main text of the article.